\documentclass[aps,prb,twocolumn,superscriptaddress,showpacs]{revtex4}

\begin{document}

\title{St\v{r}eda-like formula in spin Hall effect}
\author{Min-Fong Yang}
\email[]{mfyang@thu.edu.tw} %
\affiliation{Department of Physics, Tunghai University, Taichung,
Taiwan}
\author{Ming-Che Chang}
\email[]{changmc@phy.ntnu.edu.tw} %
\affiliation{Department of Physics, National Taiwan Normal
University, Taipei, Taiwan}
\date{\today}

\begin{abstract}
A generalized St\v{r}eda formula is derived for the spin transport
in spin-orbit coupled systems. As compared with the original
St\v{r}eda formula for charge transport, there is an extra
contribution of the spin Hall conductance whenever the spin is not
conserved. For recently studied systems with quantum spin Hall
effect in which the $z$-component spin is conserved, this extra
contribution vanishes and the quantized value of spin Hall
conductivity can be reproduced in the present approach. However,
as spin is not conserved in general, this extra contribution can
not be neglected, and the quantization is not exact.
\end{abstract}

\maketitle

Intrinsic spin Hall effect (SHE) offers new possibility of
designing semiconductor spintronic devices that do not require
ferromagnetic elements or external magnetic fields. This effect
has been theoretically predicted both in $p$-doped semiconductors
with Luttinger type of spin-orbit (SO) coupling and in $n$-doped
semiconductors with Rashba type of SO
coupling.~\cite{MNZ,MNZ1,Sinova} In the hole doped
case,~\cite{MNZ,MNZ1} the transverse spin current is generated by
the Berry curvature correction to the group velocity of a Bloch
wave packet,~\cite{Niu,Culcer} which is similar to the case of the
charge current in quantum Hall effect.~\cite{QHE} Later it is
pointed out that the intrinsic SHE can even exist in band
insulators with SO coupling, which are called spin Hall
insulators.~\cite{SHI} These spin Hall insulators would allow spin
currents to be generated without dissipation. This dissipationless
character of the intrinsic SHE again finds analogy in the quantum
Hall effect.

Pioneered by the work of Kane and Mele,~\cite{KM1} several
specific single-particle Hamiltonians
(graphene~\cite{KM1,KM2,sheng2005} and
semiconductors~\cite{bernevig2005,QWZ,ON}) giving rise to the
quantum SHE (QSHE) have been proposed, in which the intrinsic spin
Hall conductance can be quantized in units of $e/2\pi$. These
models can be considered as multiple copies of the charge Hall
effect with different values of the spin, arranged so that the
time-reversal symmetry is unbroken and the spin current is nonzero
in the presence of an applied electric field. The investigations
on its stability with respect to interactions and disorders have
just begun.~\cite{sheng2005,WBZ,XM}

Because there exists similarities between SHE and quantum Hall
effect, one may wonder if some concepts used in quantum Hall
effect can be generalized to SHE. It is well known that (integer)
quantum Hall effect can be analyzed by the St\v{r}eda
formula.~\cite{streda82} St\v{r}eda showed that, if the Fermi
level falls within the energy gap, the Hall conductance can be
given by the charge-density response to a magnetic field (from
orbital, rather than Zeeman coupling). This formula has been used
to calculate the conductance of an electron gas in the presence of
an additional periodic potential.~\cite{2D_Hofstadter} It has also
been generalized to three dimensional
systems.~\cite{3D_Hofstadter} Since the St\v{r}eda formula is
useful in quantum Hall effect, one may expect that similar formula
for spin transport may be of some help in the study of SHE. Thus
it is worthwhile to explore such a generalization.~\cite{BH}

In the present work, we show explicitly that, similar to the
well-known result of St\v{r}eda,~\cite{streda82} there are two
terms in the (static) spin Hall conductance $\sigma_{sH}$, one of
which, $\sigma_{sH}^I$, is due to the electron states at the Fermi
energy, and the other one, $\sigma_{sH}^{II}$, is formally related
to the contribution of all occupied electron states below the
Fermi energy. Since the proposed models for QSHE thus far are all
band insulators, where the density of states at the Fermi level is
zero, therefore $\sigma_{sH}^I =0$. Hence we are mainly interested
in the contribution to QSHE from $\sigma_{sH}^{II}$. Furthermore,
we show that $\sigma_{sH}^{II}$ can be separated into a conserved
part and a non-conserved part, in which the conserved part gives
rise to a St\v{r}eda-like contribution. That is, instead of
directly calculating the spin-current response to an electric
field, we can calculate the {\it spin-density} response to a
magnetic field (again from orbital, rather than Zeeman coupling)
to obtain the conserved part of $\sigma_{sH}^{II}$. However,
another contribution to $\sigma_{sH}^{II}$, which comes from the
non-conservation of spin, is not zero in general. We note that our
derivation is model independent, therefore it applies to any QSHE
model. When this general formula is applied to the aforementioned
specific models of QSHE with conserved electron spin, the
quantized values of $\sigma_{sH}$ can be easily reproduced. As the
spin in spin-orbit coupled system is not conserved in general, the
non-St\v{r}eda-like contribution in $\sigma_{sH}^{II}$ can not be
neglected, and it can result in deviation from the quantized
values, as shown in the numerical work of
Ref.~\onlinecite{sheng2005}.

Using the linear response theory, the static spin Hall
conductivity, for a $z$-component spin current flowing along the
$y$ direction under an electric field in the $x$ direction, can be
expressed as~\cite{Bastin71}
\begin{eqnarray}\label{bastin}
 \sigma_{sH}&=& i\hbar \Omega \int d\varepsilon
  f(\varepsilon)  \mbox{\rm Tr}\left[
  j_y^z \frac{dG^+(\varepsilon)}{d\varepsilon} j_x \delta(\varepsilon-H)
  \right.\nonumber \\
 &&-\left.
 j_y^z \delta(\varepsilon-H) j_x \frac{dG^-(\varepsilon)}{d\varepsilon}
 \right] ,
\end{eqnarray}
where $\Omega$ is the volume of the system, $G^{\pm}(\varepsilon)
= \lim_{\eta\rightarrow0^+} (\varepsilon-H\pm i\eta)^{-1}$ is the
operator Green function of electrons described by the Hamiltonian
$H$. $j_\mu$ ($j_\mu^\alpha$) is the charge ($\alpha$-component
spin) current density operator in $\mu$ direction. The trace goes
over every eigenstate in the space of the Hamiltonian $H$. In
Eq.~(\ref{bastin}), the delta function can be written in terms of
Green functions,
$\delta(\varepsilon-H)=-\left[G^+(\varepsilon)-G^-(\varepsilon)\right]
/ 2\pi i$.

If we keep half of Eq.~(\ref{bastin}) and make an integration by
parts on the second half, then $\sigma_{sH}$ can be expressed as
\begin{eqnarray}
\sigma_{sH}&=&\sigma_{sH}^I+\sigma_{sH}^{II} ,  \\
\sigma_{sH}^I &=& -\frac{i\hbar\Omega}{2}
  \int d\varepsilon
   \frac{\partial f(\varepsilon )}{\partial \varepsilon }
   \mbox{\rm Tr}\left[
  j_y^z G^+(\varepsilon) j_x \delta(\varepsilon-H)
  \right. \nonumber \\
 &&  \left.
  - j_y^z \delta(\varepsilon-H) j_x G^-(\varepsilon)
  \right]   , \label{eq:streda1} \\
\sigma_{sH}^{II} &=& \frac{\hbar\Omega}{4\pi}
  \int d\varepsilon f(\varepsilon)
  \mbox{\rm Tr}\left[
  j_y^z\frac{dG^-(\varepsilon)}{d\varepsilon}j_x G^-(\varepsilon)
   \right. \nonumber \\
 && -j_y^z G^-(\varepsilon)j_x \frac{dG^-(\varepsilon)}{d\varepsilon}
    -j_y^z \frac{dG^+(\varepsilon)}{d\varepsilon} j_x G^+(\varepsilon) \nonumber \\
 && \left.
    + j_y^z G^+(\varepsilon)j_x \frac{dG^+(\varepsilon)}{d\varepsilon}
   \right]  . \label{eq:streda2}
\end{eqnarray}
Here $\sigma_{sH}^I$ stems from the contribution of electrons at
the Fermi surface and $\sigma_{sH}^{II}$ formally contains the
contribution of all filled states below the Fermi energy. Note
that, at zero temperature, $\sigma_{sH}^I$ is proportional to the
density of states at the Fermi energy. Therefore, for band
insulators, where the Fermi level lies within the band gap such
that the density of states at the Fermi level is zero,
$\sigma_{sH}^I =0$. Since we are interested in the band insulators
with QSHE, we need only consider the contribution from
$\sigma_{sH}^{II}$.

In order to simplify $\sigma_{sH}^{II}$, explicit expressions of
$j_x$ and $j^z_y$ are needed. The charge current density operator
is $j_x=-e v_x/\Omega$. However, in studying SHE, the
spin is in general not conserved due to the presence of SO
interaction, thus the definition of the spin current is not unique.
Here the conventional definition for spin current is
used, that is, $j^z_y \equiv \{v_y, s_z\}/2\Omega$.

Using the cyclic property of the trace operation and the relations
\begin{eqnarray}
  && \frac{dG^\pm(\varepsilon)}{d\varepsilon}
  = - \left[G^{\pm}(\varepsilon)\right]^2  ,
  \nonumber \\
  &&i\hbar v_\mu = [r_\mu,H]
  = -\left[ r_\mu, \frac{1}{G^\pm(\varepsilon)} \right]  , \nonumber
\end{eqnarray}
we find that the spin Hall conductance $\sigma_{sH}^{II}$ can be
separated into a ``conserved" part $\sigma_{sH}^{II,(c)}$ and a
``non-conserved" part $\sigma_{sH}^{II,(n)}$,
\begin{widetext}
\begin{eqnarray}
\sigma_{sH}^{II}&=& \sigma_{sH}^{II,(c)} + \sigma_{sH}^{II,(n)} , \\
\sigma_{sH}^{II,(c)}&=&
  \frac{e}{4\pi i\;\Omega}
  \int d\varepsilon  f(\varepsilon)
  \mbox{\rm Tr}  \left\{
  s_z G^+ (x v_y - y v_x) G^+
  - s_z G^- (x v_y - y v_x) G^-
  \right\} ,   \label{streda2(c)} \\
\sigma_{sH}^{II,(n)}&=&
  -\frac{i\hbar^2 e}{4\pi \Omega}
  \int d\varepsilon  f(\varepsilon)
  \mbox{\rm Tr} \left\{
  \dot{s}_z \left[
  \left(G^+ \right)^2 v_x G^+ v_y G^+
  + G^+ v_y G^+ v_x \left(G^+ \right)^2 -\hbox{\rm H. c.} \right]
  \right\}  \nonumber \\
  && - \frac{\hbar e}{8\pi \Omega}
  \int d\varepsilon  f(\varepsilon)
  \mbox{\rm Tr} \left\{
  \dot{s}_z \left[ G^+ [y, v_x]  \left(G^+ \right)^2
  + \left(G^+ \right)^2 [y, v_x] G^+ +\hbox{\rm H. c.}  \right]
  \right\} .
\label{streda2(n)}
\end{eqnarray}

Here we call $\sigma_{sH}^{II,(n)}$ as the ``non-conserved" part
because it contains terms with $\dot{s}_z \equiv [s_z, H]/i\hbar$
and therefore vanishes if $s_z$ is conserved. By employing the
identity,~\cite{note1}
\begin{eqnarray}
     \frac{\partial}{\partial B} \mbox{\rm Tr} \left\{ s_z
     \delta(\varepsilon-H)\right\}
     =\frac{- e}{4\pi i} \mbox{\rm Tr}  \left\{
     s_z G^+ (x v_y - y v_x) G^+
    - s_z G^- (x v_y - y v_x) G^-
    \right\} , \label{identity}
\end{eqnarray}
\end{widetext}
the conserved part $\sigma_{sH}^{II,(c)}$ can be rewritten as a
generalized St\v{r}eda formula for SHE,
\begin{equation}\label{streda}
\sigma_{sH}^{II,(c)} = - \left. \frac{\partial S_z}{\partial B}
\right|_{\mu,T} ,
\end{equation}
where $B$ is the magnitude of a uniform external magnetic field in
the $z$ direction, $\mu$ is the chemical potential, $T$ is the
temperature, and $S_z$ is the $z$-component spin density,
\begin{equation}
  S_z = \frac{1}{\Omega} \int d\varepsilon
  f(\varepsilon)   \mbox{\rm Tr} \left[ s_z
  \delta(\varepsilon-H)\right]   .  \label{S_z}
\end{equation}
Eq.~(\ref{streda}) shows the relation between the spin Hall
conductivity and the derivative of the $z$-component spin density
with respect to the perpendicular magnetic field $B$.
Eqs.~(\ref{eq:streda1}), (\ref{streda2(n)}), and (\ref{streda})
are the main results of this paper.

Some comments are in order. Contrary to the case of the original
St\v{r}eda formula for charge transport,~\cite{streda82} there
always exists an extra contribution for $\sigma_{sH}^{II}$ besides
the St\v{r}eda-like one as long as the spin is not conserved.
Moreover, following the same reasoning of the present derivation,
it is obvious that, for a current carrying some {\it conserved}
quantity, there will always be a corresponding St\v{r}eda-like
formula. For example, when the total angular momentum $J_z$
consisting of both spin and orbit angular momenta is conserved,
the Hall conductivity for the {\it total-angular-momentum current}
can be calculated through the {\it total-angular-momentum density}
response to a magnetic field. That is, a St\v{r}eda-like formula
in which $s_z$ is replaced by $J_z$ can be derived.

In the following, we illustrate the application of the present
approach to some of the aforementioned QSHE models. As mentioned
earlier, since the proposed models for QSHE thus far are all band
insulators, thus what we need to consider is just the contribution
from $\sigma_{sH}^{II}$. It is shown in
Ref.~\onlinecite{bernevig2005} that, in the presence of a linear
strain gradient, degenerate quantum Landau levels can be created
by the SO coupling in conventional semiconductors, and then
$\sigma_{sH}$ becomes quantized in units of $e/2\pi$. This result
can be understood as follows. Even though the intrinsic SO
coupling is present, $s_z$ remains conserved in this case.
Therefore, the non-conserved part $\sigma_{sH}^{II,(n)} =0$ and
the spin Hall conductance comes from the St\v{r}eda-like one
[Eq.~(\ref{streda})] only. To calculate the spin-density response
to an external magnetic field $B$ in the $z$ direction, we notice
that, at $B=0$, opposite effective orbital magnetic fields $\pm
B_{\rm eff}$ acting on spin $\uparrow$ and $\downarrow$ electrons
are induced through this special position-dependent SO
interaction. When an external magnetic field $B$ is turned on, the
total effective fields acting on spin $\uparrow$ and $\downarrow$
electrons will become $\pm (B_{\rm eff} \pm B)$. Therefore, for a
fixed chemical potential $\mu$, the number density of electrons
with spin $\uparrow$ and $\downarrow$ becomes $N_{\uparrow /
\downarrow} = e (B_{\rm eff} \pm B)/h$. Thus, using
Eq.~(\ref{streda}), $\sigma_{sH}=\sigma_{sH}^{II,(c)}=e/2\pi$ can
be obtained.

As the next application of our formula, we consider the QSHE model
proposed in Ref.~\onlinecite{KM1}, where the intrinsic SO coupling
in a single-layer graphene film gives rise to QSHE with
$\sigma_{sH}$ being quantized in units of $e/2\pi$. This result
can also be obtained by the present approach. For this model of
graphene {\it without} the Rashba SO coupling, spin $s_z$ is
conserved. Again, $\sigma_{sH}^{II,(n)} =0$ and the spin Hall
conductance comes from the St\v{r}eda-like one
[Eq.~(\ref{streda})] only. Because each independent subsystem of
spin direction ($\uparrow$ or $\downarrow$) is equivalent to
Haldane's spinless model,~\cite{Haldane88} following the same
reasoning provided by Haldane, an extra {\it field dependent}
number density of electrons for spin $\uparrow$ and $\downarrow$
will be $\Delta N_{\uparrow / \downarrow} = \pm e B/h$ when an
external magnetic field $B$ is turned on. Thus, using
Eq.~(\ref{streda}), $\sigma_{sH}=\sigma_{sH}^{II,(c)}=e/2\pi$ can
again be reached.

In the generic cases, spin is not conserved in the presence of SO
interaction and therefore spin current is not well-defined. As
done in Ref.~\onlinecite{MNZ1}, one can separate the spin operator
$s_z$ into a conserved part $s_z^{(c)}$ and a non-conserved part
$s_z^{(n)}$: $s_z=s_z^{(c)}+ s_z^{(n)}$, where the conserved part
$s_z^{(c)}$ consists of only intraband matrix elements of the
spin. Thus a conserved spin current can be defined by substituting
the conserved part $s_z^{(c)}$ into the conventional expression of
spin current. By employing this definition of spin current, it can
be shown that the non-conserved part $\sigma_{sH}^{II,(n)}$
vanishes and the conserved part $\sigma_{sH}^{II,(c)}$ again obeys
the St\v{r}eda-like formula, Eq.~(\ref{streda}), in which the spin
operator $s_z$ is replaced by $s_z^{(c)}$. In
Ref.~\onlinecite{PZhang}, another definition of the spin current
is proposed which includes the torque dipole density. If we use
this definition in the derivation for $\sigma_{sH}^{II}$, the
St\v{r}eda-like formula Eq.~(\ref{streda}) for the conserved part
$\sigma_{sH}^{II,(c)}$ remains valid, and the non-conserved part
$\sigma_{sH}^{II,(n)}$ remains nonzero and has
different expression from Eq.~(\ref{streda2(n)}).

To conclude, a generalized St\v{r}eda formula is derived for spin
transport, and the applications of the present approach to the
recently proposed QSHE systems are discussed. Since the original
St\v{r}eda formula has succeeded in analyzing the quantum Hall
effect in the presence a periodic potential if the Fermi level
lies within a miniband gap,~\cite{2D_Hofstadter,3D_Hofstadter} we
expect the generalized St\v{r}eda formula will be of help for the
spin transport under the same situation.~\cite{note2}
Investigation along this direction is in progress.

M.F.Y. and M.C.C acknowledge the support by the National Science
Council of Taiwan under the contracts NSC 94-2112-M-029-008 and
NSC 94-2112-M-003-017, respectively.


\begin{thebibliography}{99}
\bibitem{note1}
This identity is valid for arbitrary choices of vector potential
$\vec{A}$, nevertheless the derivation would be simpler if a
symmetric gauge that centers at the origin is used, that is,
$\vec{A} \equiv (-y, x, 0)B/2$.
\end{thebibliography}
\end{document}